\documentclass[conference]{IEEEtran}
\IEEEoverridecommandlockouts

\usepackage{cite}
\usepackage{amsmath,amssymb,amsfonts}
\usepackage{algorithmic}
\usepackage{graphicx}
\usepackage{textcomp}
\usepackage{xcolor}
\usepackage{url}
\usepackage[T1]{fontenc}
\usepackage[english]{babel}
\usepackage{array}
\usepackage{stfloats}
\usepackage{units}
\usepackage{cite}

\def\BibTeX{{\rm B\kern-.05em{\sc i\kern-.025em b}\kern-.08em
		T\kern-.1667em\lower.7ex\hbox{E}\kern-.125emX}}

\addto\captionsenglish{}

\begin{document}

\title{Analysis of Cooperative Hybrid ARQ with Adaptive Modulation and Coding on a Correlated Fading Channel}

\author{\IEEEauthorblockN{Ibrahim Ozkan, \textit{Student Member, IEEE}}
\IEEEauthorblockA{Department of Electrical and Electronics Engineering\\
Hacettepe University, Ankara, Turkey 06800\\
e-mail: ozkan@ee.hacettepe.edu.tr}
}

\maketitle

\begin{abstract}
	In this study, a cross-layer design which combines adaptive modulation and coding (AMC) and hybrid automatic repeat request (HARQ) techniques for a cooperative wireless network is  investigated analytically. Previous analyses of such systems in the literature are confined to the case where the fading channel is independent at each retransmission, which can be unrealistic unless the channel is varying very fast. On the other hand, temporal channel correlation can have a significant impact on the performance of HARQ systems. In this study, utilizing a Markov channel model which accounts for the temporal correlation, the performance of non-cooperative and cooperative networks are investigated  in terms of packet loss rate and throughput metrics for Chase combining HARQ strategy.
\end{abstract}


\section{Introduction}
In wireless networks, severe channel effects of signal fading arising from multi-path propagation can be  mitigated through the use of diversity.  Multiple antenna solutions are well-known methods that mitigate the effects of fading in wireless environments. However, in cases where multiple antenna solutions are not feasible due to numerous limitations, it is crucial to resort to other solutions, such as adapting to time varying channel and cooperation diversity.

Two well-established methods to enhance the system performance by adapting to the time-varying wireless channel are AMC  at the physical layer and HARQ at the data link and physical layers. AMC improves the system performance by selecting a suitable code and modulation pair regarding the time-varying channel behavior. HARQ is an error-correction technique that combines multiple ARQ retransmissions, increasing the number of retransmissions for worse channel conditions.

Implementing AMC and HARQ  in a cooperative network is a deeply cross-layer and sensible approach which can achieve spatial diversity gain of user cooperation as well as adapt the time-varying nature of wireless channels. It is apparent that the performance of the strategy of combining cooperation, AMC, and HARQ would be dependent on channel correlation in time. The goal in this study is to investigate the performance of HARQ with AMC in both cooperative and non-cooperative scenarios under a correlated channel model.

The performance of HARQ was analyzed  using the criteria of packet error rate and spectral efficiency, which we shall also use in this study \cite{nc_harq_arq1,nc_harq2}. In another study, the performance analysis of HARQ system combined with AMC is presented \cite{nc_harq1}. The channel is modeled as independent fading for each packet transmission in \cite{nc_harq_arq1,nc_harq2,nc_harq1}. Channel fading correlated in time is adopted in \cite{nc_harq_corr1}, where the performance of HARQ is studied. 

A cooperative rate adaptive wireless network with HARQ is investigated on independent fading channels by adopting capacity theorem approach in \cite{cc_ir3, it_coop2}. Authors propose an improved relaying protocol by combining various relaying strategies in \cite{coop_harq_arp} for a cooperative network which consists of a single relay without employing AMC scheme.

Performance of HARQ with AMC in a cooperative network is analyzed in \cite{amc_harq1,amc_harq2}. In these papers authors adopt a quasi-static Rayleigh fading model with a correlation between retransmissions. Authors of \cite{harq_add} analyze performance of HARQ protocol on a cooperative network from an information theoretical perspective to reveal the optimum transmission rate under time-correlated fading conditions. In a work that is perhaps the most closely related to this study, in \cite{amc_thresholds2} authors analyze the performance of HARQ with AMC in a cooperative network under correlated fading. The difference in \cite{amc_thresholds2} and this work is the type of HARQ technique adopted. Furthermore, this study can be considered as an extension of the framework that is presented in \cite{selfie}. 

Furthermore, as emphasized in \cite{vangelista2018performance} recently, for Internet of things (IoT) or more specifically machine-to-machine (M2M) communication protocols, cooperative communication and HARQ are promising techniques for future networks. The main contribution of this study is the analytic expression of throughput and packet loss rate for both cooperative and non-cooperative Chase combining HARQ protocols with AMC scheme employed in time-correlated Rayleigh fading environment. 

The rest of the study is organized as follows. The system and the channel model will be presented in Section II. Also the cross-layer design approach will be discussed in this section. Subsequently, the analytical solutions for throughput and packet loss rate are introduced in Section III. The numerical results of the analyses and simulations will be exhibited in Section IV before a brief conclusion in Section V.

\section{SYSTEM MODEL FOR CHASE COMBINING HARQ WITH AMC}
\subsection{System Description}
A wireless communication link between a single-antenna source (S) node and a destination (D) node employs AMC at link layer and a HARQ protocol at Data-Link Layer (DLL) is to be studied.

\begin{figure}[b!]
	\centering
	\includegraphics[width=0.47\textwidth, height = 0.47\textheight,keepaspectratio]{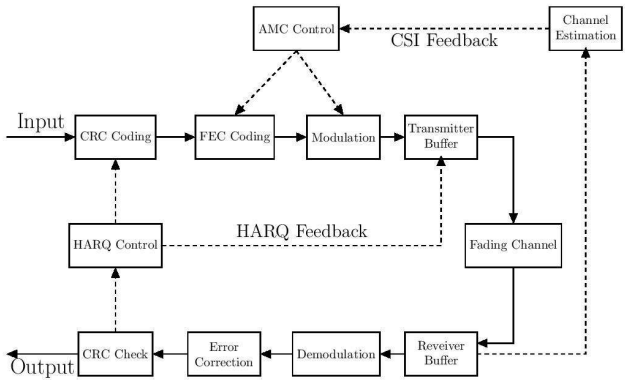}
	\caption{Depiction of cross-layer design of a wireless network.}
	\label{fig:cross}
\end{figure}

In the cross-layer model of concern, AMC Controller selects the proper modulation and forward-error correction (FEC) coding scheme pair regarding to the channel state information (CSI) which is assumed as a perfect estimation of the fading channel and fed back reliably by the receiver on a frame-by-frame basis.  A detailed structure can be seen in Fig. \ref{fig:cross}.

If a packet is decided to be erroneous, the receiver feeds a negative acknowledgment (NACK) via HARQ controller to request a retransmission. The receiver does not discard the erroneously received packet and keeps it in the buffer to combine with the succeeding transmission(s) by means of maximal-ratio combining (MRC) and tries to obtain an error-free packet. The retransmissions of a packet are performed until the maximum number of transmissions, constrained by the system requirements is reached. If a packet is not obtained error-free after the maximum allowed transmissions, it is declared to be lost. Briefly, this implies a truncated type Chase combining HARQ protocol \cite{trunc}.

\subsection{Channel Model and AMC Scheme}
It is assumed a frequency flat Rayleigh fading model with a stationary and ergodic gain and zero-mean additive white Gaussian noise. The channel gain is assumed to remain constant during the entire time of a packet transmission, but it varies from one usage to another based on a time-correlated process. The range of the channel SNR is divided into $N_S+1$ non-overlapping consecutive intervals, and each is denoted by $[\Gamma_n, \Gamma_{n+1}),\; n = 0, 1,\cdots, N_S$ where $\Gamma_0 = 0 $ and $\Gamma_{N_S+1} = \infty$. If the channel SNR is in the $n$-th interval, then the channel is in State $n$. A Markov channel model is considered where it is defined a discrete-time Markov process $\Psi_j$, with $\Psi_j = n$ represents the channel state at time interval $j$ being in State $n$. The channel state transition probability matrix $\mathbf{P}$ has dimension
$(N_S+1) \times (N_S+1)$, where  $(k,n)$-th element is \cite{CCtan}
\begin{eqnarray}
\nonumber
P_{n,k}^{\tau_f^{n}} &=&\left. P\left\{\Psi_{j + 1} = k \right|  \Psi_j = n\right\}\\
&=& \frac{1}{\pi_n} \int_{\displaystyle\zeta_n}^{\displaystyle\zeta_{n+1}} \int_{\displaystyle\zeta_k}^{\displaystyle\zeta_{k+1}} p_{r_1, r_2}(r_1,r_2;\rho_n) \,\text{d}r_1 \text{d}r_2,
\label{eq:transprob}
\end{eqnarray}
where $\tau_f^{n}$ is the time interval between two successive transmissions when AMC mode $n$ is chosen 
with probability $\pi_n$, $\zeta_i = \sqrt{\nicefrac{\Gamma_i}{\bar{\gamma}}}$, $\bar{\gamma}$ is the mean SNR and $p_{r_1, r_2}(r_1,r_2;\rho_n)$ is the bivariate Rayleigh joint probability density function (pdf) of two envelopes $r_1$ and $r_2$:
\begin{equation}
p_{r_1, r_2}(r_1,r_2;\rho_n) = \frac{4 r_1 r_2 e^{ -\frac{ \left(r_1^2 + r_2^2 \right) }{1-\rho_n}} }{1-\rho_n}  I_0 \left(\frac{2 r_1 r_2\sqrt{\rho_n}}{1-\rho_n}\right).
\label{eq:bivpdf}
\end{equation}
In (\ref{eq:bivpdf}), $\rho_n$ is the envelope correlation coefficient for Rayleigh fading process between the SNR of two consecutive transmissions, $I_0(\cdot)$ is zeroth order modified Bessel function and expressed as follows
\begin{equation}
\rho_n = J_0^2(2\pi f_D \tau_f^{n}),
\end{equation} where $f_D$ is the maximum Doppler frequency and  $J_0(\cdot)$ is zeroth order Bessel functions.
The integral equation in (\ref{eq:transprob}) can be numerically calculated by following the method proposed in \cite{simon}.
\subsection{Cross-Layer Design of HARQ}
Since some finite delay can be affordable for many communications systems, the maximum allowable retransmission number, $N_R$, has to be limited for an individual packet. On the other hand, to maintain a favorable data flow, the packet error rate at each transmission must be guaranteed to be below a certain level $P_\epsilon$. This constraint on error rate is usually set by the  quality of service requirements and may be varied from one system to another. The delay and packet error rate requirements impose constraints on HARQ at the data link layer \cite{per_approx}. AMC scheme design will be accomplished after setting the thresholds of the channel states. A distinct transmission mode will be assigned to any channel states while satisfying packet error rate $P_\epsilon$ and a instantaneous signal-to-noise ratio (SNR) of the channel. The parameters $a_m$, $g_m$ and $\gamma_{pm}$ are mode dependent constants and used for the packet error event modeling of the convolutionally coded packets and can be found in Table \ref{tab:amctable}.
\begin{equation}
\text{PER}_m(\gamma)=\begin{cases}
1, & \text{if $0 \leq \gamma < \gamma_{pm}$},\\
a_m e^{-g_m\gamma}, & \text{if $\gamma \geq \gamma_{pm}$}
\end{cases}
\label{eq:per_aprox}
\end{equation}

Let $\overline{\text{PER}}_{m}$ represent the average packet error rate of the first transmission of a packet when mode $m$ is chosen. Thus, $\overline{\text{PER}}_{(m)}$  can be expressed as
\begin{equation}
\overline{\text{PER}}_{m}  =\frac{1}{\pi_m} \int_{\Gamma_m}^{\Gamma_{m+1}} a_m \exp(-g_m\gamma) p_{\gamma}(\gamma) \, \text{d} \gamma \leq P_\epsilon,
\label{eq:pertgt}
\end{equation}
The thresholds are obtained by solving  (\ref{eq:pertgt}) recursively for each AMC mode $m$. Since the retransmissions for an individual packet are combined by means of MRC at the receiver side, the received SNR is accumulated and the average packet error rate after each retransmission is guaranteed to be below $P_\epsilon$ as suggested in \cite{amc_thresholds2}.

\begin{table*}[t!]
	\caption{Transmission modes specifications and PER model parameters \cite{hiperlan2}}
	\label{tab:amctable}
	\centering
	\scalebox{1.5}[1.5]{
		\begin{tabular}{c || c | c | c | c | c | c }
			\hline
			\textbf{Mode $m$} & $\mathbf{1}$ &$\mathbf{2}$  & $\mathbf{3}$  &$\mathbf{4}$  & $\mathbf{5}$  & $\mathbf{6}$ \\
			\hline
			\hline
			\textit{Modulation} & \text{BPSK} & \text{QPSK} &\text{QPSK} & \text{16-QAM}& \text{16-QAM} & \text{64-QAM}\\
			\hline
			\textit{Coding rate} & $1/2$ & $1/2$  & $3/4$ & $9/16$ & $3/4$ & $3/4$ \\
			\hline
			\textit{Rate: $R_m$} bps& $0.50$& $1.00$ & $1.50$ & $2.25$ & $3.00$ & $4.50$\\
			\hline
			$a_m$ & $274.7229$ & $90.2514$ & $67.6181$ & $50.1222$ & $53.3987$ & $35.3508$  \\
			\hline
			$g_m$ & $7.9932$ & $3.4998$ & $1.6883$ & $0.6644$ & $0.3756$& $0.0900$  \\
			\hline
			$\gamma_{pm} (dB)$ & $-1.5331$ & $1.0942$ & $3.9722$ & $7.7021$ & $10.2488$ & $15.9784$  \\
			\hline
	\end{tabular}}
\end{table*}

\subsection{Cooperative network}
\begin{figure}[b!]
	\centering
	\includegraphics[width=0.4\textwidth, height = 0.4\textheight,keepaspectratio]{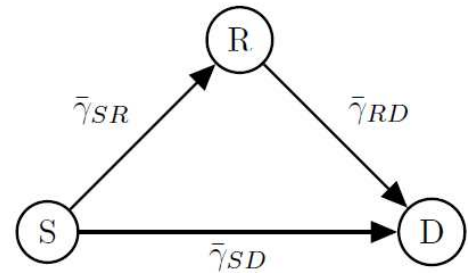}
	\caption{Figurative representation of a cooperative network.}
	\label{fig:sdr}
\end{figure}

The cooperative transmission can be divided into two fundamental phases. In the first phase, node S transmits the packet after assigning an AMC mode regarding to the state of the link that ties it to node D. In this phase, both node D and the relay (R) try to decode the packet. If node D receives an error-free packet, it broadcasts an ACK message and transmission of new packet is started by node S. Otherwise node D broadcasts a NACK message and the second phase starts. A figurative depiction of cooperative network is given in Fig. \ref{fig:sdr}.

In the second phase, cooperation is triggered if an ACK is emitted by node R. The retransmission is performed by the node R if it was able to decode the initial transmission successfully. Node R retransmits the packet based on the same AMC mode selected by the node S  since the packets are combined by means of MRC. If node R has not obtained the packet error-free, node S  will perform the retransmission and the cooperative network reduces to non-cooperative network.

Assuming only one retransmission ($N_R=1$) is allowed for each packet, if a packet is not decoded error-free after two transmissions, packet is dropped and node S skips to transmission of a new packet.

In cooperative network, each link is subjected to statistically independent small-scale flat Rayleigh fading and large-scale path loss. 

\section{THROUGHPUT AND PACKET LOSS RATE  ANALYSIS}
The packet loss rate (PLR) is defined as average ratio of the lost packets in the network, and throughput ($ \eta$) as the average ratio of successfully delivered packets to the number of all transmissions performed in the network. We first perform the analysis for a non-cooperative network without node R and then extend the results to cooperative network.
\subsection{Non-cooperative Network}
\subsubsection{Packet Loss Rate}
Average probability of a packet to be lost in non-cooperative network can be defined as the joint probability of the decoding error events after the first and the second transmissions of the packet. That is,
\begin{equation}
\text{PLR}_{NC} = \text{P} \{F_{SD}^1, F_{SD}^2\}
\label{eq:plr1}
\end{equation} where $F_{SD}^j$ is the decoding error event at node D  after the $j$-th transmission and it can be rewritten as $F_{SD}^j = P\{S_{SD}^1, S_{SD}^2, \cdots, S_{SD}^j\}$ in order to emphasize the dependence of a decoding event to the previous transmissions of a packet due to use of MRC at the receiver where $S_{SD}^j$ is the state of $SD$ link during the $j$-th transmission.

The packet loss rate for non-cooperative network is 
\begin{equation}
\text{PLR}_{NC} = \hspace{-0.2cm}\sum_{S_{SD}^1 = 1}^{N_S} \sum_{S_{SD}^2 = 1}^{N_S} \hspace{-0.2cm}P\{S_{SD}^1,S_{SD}^2\} 
P \{F_{SD}^1,F_{SD}^2 | S_{SD}^1,S_{SD}^2 \}
\label{eq:plr2}
\end{equation}
By considering following identities that stem from Markov process, (\ref{eq:plr2}) can further be expanded
\begin{equation}
P\{S_{SD}^1,S_{SD}^2\} = P\{S_{SD}^{2}|S_{SD}^{1}\}P\{S_{SD}^1\},
\label{eq:probfail}
\end{equation}
\begin{equation}
P\{F_{SD}^1,F_{SD}^2 | S_{SD}^1,S_{SD}^2 \} =
P \{F_{SD}^1|S_{SD}^1\} P\{F_{SD}^2|S_{SD}^2,S_{SD}^1\}
\label{eq:fail}
\end{equation}
The average probability decoding error event after the first transmission can be expressed as
\begin{equation}
P \{F_{SD}^1|n\} = \int \text{PER}_n(\gamma^1) p_{\gamma^1}(\gamma^1|n ) \, \text{d}\gamma^1 .
\label{eq:err1nc}
\end{equation}
The retransmission of the packet will be performed $\tau_f^{n}$ seconds after the first transmission in a new channel state but the AMC mode will be kept unchanged so that Chase combining can be performed.  The average probability of decoding error when retransmission of the packet is performed in state $S_{SD}^2=k$, combined with the first transmission is expressed as
\begin{equation}
P\{F_{SD}^2|n,k\} = \int  \text{PER}_n(\gamma_o) p_{\gamma_o}(\gamma_o|n,k) \, \text{d}\gamma_o ,
\label{eq:err2nc}
\end{equation} where $p_{\gamma_o}(\gamma_o)$ represents the pdf of the sum of two correlated SNR's of the first transmission and the retransmission \cite{mrc2}.
\subsubsection{Throughput}
The throughput for the non-cooperative case is found as
\begin{equation}
\eta_{NC} = \frac{1- \text{PLR}_{NC}}{ (N_R +1)\text{PLR}_{NC} + \sum_{N_T = 1}^{(N_R + 1)} N_T P_{NC}^{N_T}},
\label{eq:thrnc}
\end{equation}
where $N_T$ is the number of transmissions performed to deliver a packet error-free and its average probability $P_{SD}^{N_T}$ is expressed as the joint probability of 
\begin{equation}
P_{NC}^{N_T}=	P\{F_{SD}^1, F_{SD}^2\,\cdots,\overline{F_{SD}^{N_T}} \}
\label{eq:Pnt}
\end{equation}
In (\ref{eq:Pnt}), $\overline{F_{SD}^{N_T}} $ is the event of decoding without error after exactly $N_T$-th transmission of the packet.

\subsection{Cooperative Network}
In order to express packet loss rate and throughput analytically for cooperative HARQ network, the average probability of decoding error event at node R, $P\{F_{SR}^1\}$, and the average probability error of decoding error event at node D when retransmission is performed by node R, $\text{PLR}_{SRD}$, must be stated in addition to the results obtained for the non-cooperative network in the previous section.
\subsubsection{Packet Loss Rate}
The packet loss rate for non-cooperative network can analytically be expressed as
\begin{equation}
\label{eq:plrc}
\text{PLR}_{C}= \text{PLR}_{NC}P\{F_{SR}^1\} + 
\text{PLR}_{SRD} P\{\overline{F_{SD}^1}\} ,
\end{equation}
where 
\begin{equation}
\label{eq:plrsrd}
\text{PLR}_{SRD} = \sum_{S_{SD}^1 = 1}^{N_S} \sum_{S_{RD}^2 = 1}^{N_S} P\{F_{SRD}^2 | S_{SD}^1,S_{RD}^2 \}.
\end{equation}
The probability of decoding error event when the first transmission performed by S and the second by R is
\begin{eqnarray}
\label{eq:cnk}
\nonumber	&&\hspace*{-1cm}P\{F_{SRD}^2 | S_{SD}^1 =n,S_{RD}^2 \} =  \\
&&\hspace*{-1cm}\frac{1}{\pi_n} \int\displaylimits_0^{\infty} \int\displaylimits_{\Gamma_n}^{\Gamma_{n+1}} \text{PER}_n(\gamma^1 + \gamma^2) p_{\gamma^1}(\gamma^1) p_{\gamma^2}(\gamma^2) \, \text{d}\gamma^1 \text{d}\gamma^2
\end{eqnarray}
\subsubsection{Throughput}
The throughput expression of the cooperative network is obtained in a similar way as in (\ref{eq:thrnc}) except for the probability of a packet to be delivered without error after exactly $N_T$ transmission(s). This can be expressed as
	\begin{equation} 
\resizebox{7.5cm}{0.65cm}{ $
	P_{C}^{N_T}=\begin{cases}
	P\{\overline{F_{SD}^1}\} &, N_T = 1\\[6pt]
	P\{F_{SR}^1\}P\{F_{SD}^1,\overline{F_{SD}^2} \}+
	P\{\overline{F_{SR}^1}\}P\{\overline{F_{SRD}^2}\}&, N_T = 2
	\end{cases}$}
\label{eq:pctnt}
\end{equation}
 \begin{figure}[t!]
	\centering
	\includegraphics[width=0.5\textwidth, height = 0.5\textheight,keepaspectratio]{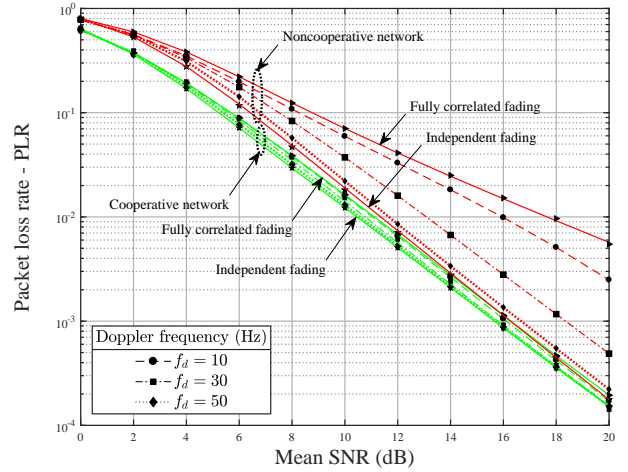}
	\caption{Packet loss rate performance. Markers show the simulation results while curves depict the analytical results.}
	\label{fig:plr}
\end{figure}

\section{RESULTS}
The numerical results are obtained by running simulations on Matlab. The longest packet duration in the system is set to be $1$ ms  with a constant round-trip-time (RTT) of $2$ ms for each mode of transmission. A non-cooperative network and a cooperative network are analyzed with packet loss constraint, $P_\epsilon = 10^{-4}$ and the maximum number of allowable retransmission, $N_R = 1$.

The numerical results are obtained for different channel correlation scenarios by adjusting the maximum Doppler frequency to  $f_D = \{10,\; 30,\; 50\}$ Hz. These Doppler frequency settings correspond to the correlation coefficients approximately $\rho =\{0.9,\; 0.8,\; 0.6\}$ between the successive transmissions. The correlation coefficients are given as approximate values because of the difference in packet duration of each transmission mode. Nevertheless, these three different Doppler settings may be considered as different situations in a highly correlated wireless channel.  Also the extreme cases such as fully correlated and independent fading results are included. For cooperative scenario, considering large-scale path loss, the relation between the mean SNR values of the links are assumed to be as $\bar\gamma_{SR} = \bar\gamma_{RD} = 4\bar\gamma_{SD}$, which is a typical scenario when node R is located around the midpoint of nodes S and D.
\begin{figure}[t!]
	\centering
	\includegraphics[width=0.5\textwidth, height = 0.5\textheight,keepaspectratio]{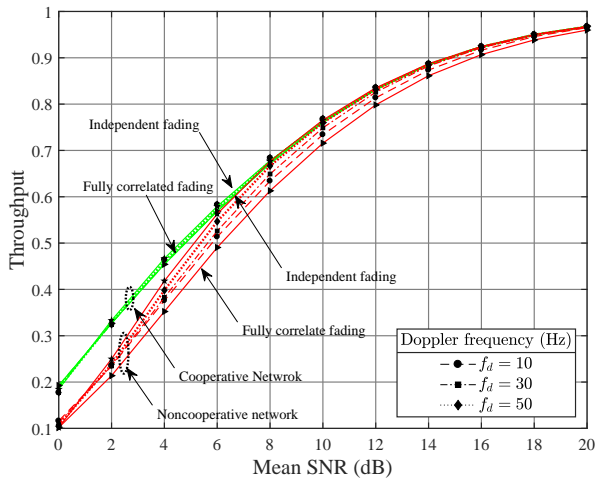}
	\caption{Throughput performance. Markers show the simulation results while curves depict the analytical results.}
	\label{fig:thr}
\end{figure}

As can be seen from Fig. \ref{fig:plr},  cooperative scenario outperforms non-cooperative scenario considering packet loss rate.  It can also be concluded that as the maximum Doppler frequency increases, the PLR of non-cooperative scenario decreases dramatically. Although a limited improvement in PLR performance of cooperative network can be observed, it is not worthwhile as is non-cooperative scenario. This stems from the fact that only one retransmission of a packet is allowed in the network, there is no chance to observe the effect of correlation in the relay links. Cooperative network can achieve substantial gain over non-cooperative one when especially the mean SNR is low, while the throughput gain reduces as the mean SNR increases which can be observed in Fig. \ref{fig:thr}.

Additionally, concerning both performance criteria, cooperative network is more robust to changes in the Doppler spread of the environment, which is a probable case in real-life applications. Inclusion of one relay can improve the performance of the network in terms of stability and reliability.
\section{ CONCLUSION}
In the future works, it is desired to analyze the cooperative network with arbitrary number of relays that are scattered arbitrarily around source and destination nodes, which may comprise more realistic cases that can be encountered in vehicle-to-everything (V2X) and IoT networks.

Restricting number of retransmissions to one ($N_R = 1$) may be considered as a weak aspect that should be improved. However, it is conceivable that in a wireless environment with multiple relays that move randomly around the source and destination nodes, it is not reasonable to expect to experience the effect of correlation between the successive transmissions. From this point of view, the restriction on number of retransmissions may be a reasonable choice.

The channel models and other parameters can be easily modified to assess the system performance in various environments in the future works.

\bibliographystyle{IEEEtran}
\bibliography{IEEEabrv,ref_ioea}

\end{document}